\documentclass[prd,preprint,tightenlines,floatfix,showpacs,preprintnumbers,nofootinbib,eqsecnum,superscriptaddress]{revtex4}

 \usepackage[dvips,final]{graphicx}
  \usepackage{amssymb}
    \usepackage{amsfonts}
     \usepackage{epsfig}
      \usepackage{bm}% bold math
\def\Pom{{\bf I\!P}}

\newcommand{\bDelta}{\mbox{\boldmath $\Delta$}}

\newcommand{\bPhi}{\mbox{\boldmath $\Phi$}}

\newcommand{\bkappa}{\mbox{\boldmath $\kappa$}}
\newcommand{\bp}{\mbox{\boldmath $p$}}

\newcommand{\br}{\mbox{\boldmath $r$}}

\newcommand{\bk}{\mbox{\boldmath $k$}}

\newcommand{\bM}{\mbox{\boldmath $M$}}

\newcommand{\half}{{1\over 2}}
\def\lsim{\mathrel{\rlap{\lower4pt\hbox{\hskip1pt$\sim$}}
    \raise1pt\hbox{$<$}}}         %less than or approx. symbol
\def\gsim{\mathrel{\rlap{\lower4pt\hbox{\hskip1pt$\sim$}}
    \raise1pt\hbox{$>$}}}         %greater than or approx. symbol

\begin{document}

\thispagestyle{empty} \preprint{\hbox{}} \vspace*{-10mm}

\title{Exclusive photoproduction of $\phi$ meson \\
in $\gamma p \to \phi p$ and $p p \to p \phi p$ reactions}

\author{A. Cisek}
\email{Anna.Cisek@ifj.edu.pl}
\affiliation{Institute of Nuclear Physics PAN, PL-31-342 Cracow,
Poland} 
\author{W. Sch\"afer}
\email{Wolfgang.Schafer@ifj.edu.pl}
\affiliation{Institute of Nuclear Physics PAN, PL-31-342 Cracow,
Poland} 
\author{A. Szczurek}
\email{Antoni.Szczurek@ifj.edu.pl}
\affiliation{Institute of Nuclear Physics PAN, PL-31-342 Cracow,
Poland} 
\affiliation{University of Rzesz\'ow, PL-35-959 Rzesz\'ow,
Poland}

\date{\today}

\begin{abstract}
The amplitude for $\gamma p \to \phi p$ is calculated 
in a pQCD $k_{T}$ - factorization approach. 
The total cross section for this process is compared with HERA data.
Total cross section, as a function of photon-proton energy and 
photon virtuality, is calculated. 
We also discuss the ratio of $\sigma_{L}$/$\sigma_{T}$ and 
the dependence on the mass of the strange quark. The amplitude for
$\gamma p \to \phi p$ is used to predict the cross section 
for exclusive photoproduction of $\phi$ meson in proton-proton 
collisions. Absorption effects are included.
The results for RHIC, Tevatron and LHC energies are presented. 
\end{abstract}

%Keywords:

\maketitle

%---------------------
\section{Introduction}
%---------------------

It was shown recently how to calculate
the cross section for exclusive production of heavy
quarkonia in the $\gamma p \to V p$ \cite{INS06,Ivanov03} as well as 
in $p p \to p V p$ \cite{SS07,RSS08} collisions within the 
$k_\perp$-factorization approach.
The same formalism was also used recently to calculate exclusive 
production of the $Z^0$ boson \cite{CSS09}. 
In the latter case the corresponding cross section is however 
very small and cannot be measured at present luminosities.

In order to describe the photoproduction process in a strict
pQCD framework, at least a hard scale must be present, which
could be for example either a quark mass for heavy quarkonia, 
or a large photon virtuality for deep inelastic diffractive 
processes. The $k_\perp$--factorization formalism described
in \cite{INS06,Ivanov03} allows, however, a smooth continuation
from the hard to soft regime, and therefore can be 
also used to model the photoproduction of light vector mesons.
This derives from the fact that the $k_\perp$--factorization
is a momentum space version of the
color--dipole approach, where the soft region is modelled
by the behaviour of the dipole cross section at large dipole
sizes \cite{Kolya_CT}.

The ZEUS collaboration at HERA has measured exclusive production
of $\phi$ mesons in the $\gamma p \to \phi p$ reaction \cite{ZEUS96}
and in $\gamma^* p \to \phi p$ reaction \cite{ZEUS05}.
Here we wish to test the results of the approach against 
the HERA data as well as to make predictions for 
the $p \bar p \to p \phi \bar p$ reaction for the Tevatron 
and the $p p \to p \phi p$ reaction for RHIC and LHC. 

\vspace{1.0cm}

%--------------------------------------------------------------
\section{Photoproduction process $\gamma p \to \phi p$}
%--------------------------------------------------------------

\begin{figure}[!h]
\includegraphics[width=0.6\textwidth]{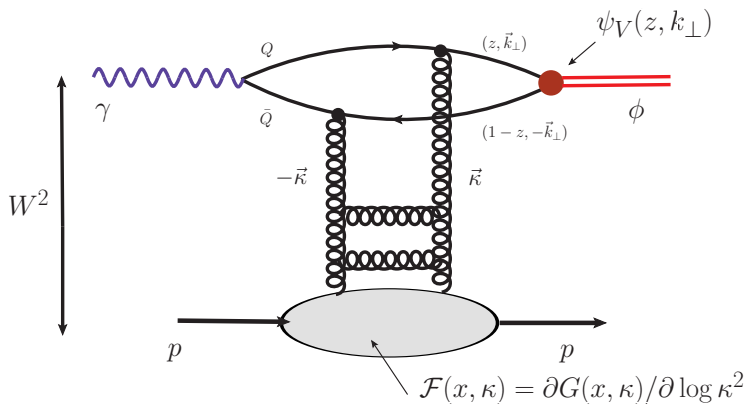}
   \caption{\label{VM_amplitude_photo}
   \small  A sketch of the amplitude for exclusive photoproduction $\gamma p \to \phi p$.}
\end{figure}

The amplitude for the reaction is shown schematically 
in Fig.\ref {VM_amplitude_photo}.
This amplitude has been derived for the first time 
in Ref.\cite{INS06},\cite{Ivanov03}. 
The imaginary part of the amplitude can be writen as:
\begin{eqnarray}
\Im m \, {\cal M} =
W^2 \frac{c_{v} \sqrt{4 \pi \alpha_{em}}}{4 \pi^2} \, 
\int 
{d\kappa^2 \over \kappa^4} \alpha_{S}(q^2)  {\cal{F}}(x_{1},x_{2},k_{1},k_{2})
\int \frac{dzd^{2}k}{z(1-z)} I(\lambda_{V},\lambda_{\gamma}) \; ,
\end{eqnarray}
where $I(\lambda_{V},\lambda_{\gamma})$ have the form:
\begin{eqnarray}
I(L,L) = 4QMz^{2}(1-z)^{2} \Big [1+ \frac{(1-2z)^{2}}{4z(1-z)}\frac{2m_{q}}{M+2m_{q}} \Big ] \psi_v(z,k) \Phi_{2}
\; ,
\end{eqnarray}
\begin{eqnarray}
I(T,T) = m_{q}^{2} \psi_v(z,k) \Phi_{2} + \Big [z^{2}+(1-z)^{2} \Big] (  \psi_v(z,k) \bk \bPhi_{1})+
\\
\nonumber
\frac {m_{q}}{M+2m_{q}} \Big [(k^{2} \psi_v(z,k)) \Phi_{2}-(2z-1)^{2}(\bk \bPhi_{1}) \psi_v(z,k) \Big ]
\; .
\end{eqnarray}
In the expressions above $\psi_v(z,k)$ is the meson light-cone wave function and 
$\bPhi_{1}$, $\Phi_{2}$ are given by (ref. \cite{INS06}):

\begin{equation}
\Phi_{2}= -{1
\over (\br+\bkappa)^2 + \varepsilon^2} -{1 \over
(\br-\bkappa)^2 + \varepsilon^2} + {1 \over (\br +
\bDelta/2)^2 + \varepsilon^2} + {1 \over (\br -
\bDelta/2)^2 + \varepsilon^2} \, ,
\end{equation}

\begin{equation}
\bPhi_{1} =
-{\br + \bkappa \over (\br+\bkappa)^2 +
\varepsilon^2} -{\br - \bkappa \over (\br-\bkappa)^2
+ \varepsilon^2} + {\br + \bDelta/2 \over (\br +
\bDelta/2)^2 + \varepsilon^2} + {\br - \bDelta/2 \over
(\br - \bDelta/2)^2 + \varepsilon^2} \,,
\end{equation}
where: $\br = \bk + (z - \frac{1}{2}) \bDelta$ and $\epsilon^{2} = 
m_{q}^{2} + z (1 - z) Q^{2}$.

\vspace{0.5cm}

The quantity ${\cal F}(x_1,x_2,\bkappa_1,\bkappa_2)$ is
the off-diagonal unintegrated gluon distribution, which
for small $\Delta^2$, within the diffraction cone 
can be approximated as:
\begin{equation}
{\cal{F}}(x_{1},x_{2},k_{1},k_{2}) =
{\cal{F}}(x_{eff},\kappa^2) \ \exp \Big(\frac{-B \bDelta^2}{2} \Big)
\, ,
\end{equation}
where
$x_{eff} = c_{s} \Big(\frac{M_{V}^2 + Q^2}{W^2} \Big), c_s$ = 0.41
and the forward unintegrated gluon distribution is 
taken from Ref.\cite{IN02}
where it was found in the analysis of the deep-inelastic
scattering data.

In the forward scattering limit, i.e.
for $\bDelta =0$ azimuthal integrations 
can be performed analytically (see \cite{RSS08}),
and one obtains the following representation for the
imaginary part of the amplitude for the transverse photons:
\begin{eqnarray}
\Im m \, {\cal M}_{T}(W,\Delta^2 = 0,Q^{2}) =
W^2 \frac{c_v \sqrt{4 \pi \alpha_{em}}}{4 \pi^2} \, 2 \, 
 \int_0^1 \frac{dz}{z(1-z)}
\int_0^\infty \pi dk^2 \psi_V(z,k^2)\\
\nonumber
\int_0^\infty
 {\pi d\kappa^2 \over \kappa^4} \alpha_S(q^2) {\cal{F}}(x_{eff},\kappa^2)
\Big( A_0(z,k^2) \; W_0(k^2,\kappa^2) 
     + A_1(z,k^2) \; W_1(k^2,\kappa^2)
\Big) \, ,
\end{eqnarray}
where
\begin{eqnarray}
A_0(z,k^2) &=& m_q^2 + \frac{k^2 m_q}{M + 2 m_q}  \, ,
\\
A_1(z,k^2) &=& \Big[ z^2 + (1-z)^2 
    - (2z-1)^2 \frac{m_q}{M + 2 m_q} \Big] \, \frac{k^2}{k^2+ \epsilon^{2}} \, ,
\end{eqnarray}
\begin{eqnarray}
W_0(k^2,\kappa^2) &=& 
{1 \over k^2 + \epsilon^2} - {1 \over \sqrt{(k^2-\epsilon^2-\kappa^2)^2 + 4 \epsilon^2 k^2}}
\, , 
\\
W_1(k^2,\kappa^2) &=& 1 - { k^2 + \epsilon^2 \over 2 k^2}
\Big( 1 + {k^2 - \epsilon^2 - \kappa^2 \over 
\sqrt{(k^2 - \epsilon^2 - \kappa^2)^2 + 4 \epsilon^2 k^2 }}
\Big) \, .
\end{eqnarray}

The imaginary part of the amplitude for longitudinal photons is given as:
\begin{eqnarray}
\Im m \, {\cal M}_{L}(W,\Delta^2 = 0,Q^{2}) =
W^2 \frac{c_v \sqrt{4 \pi \alpha_{em}}}{4 \pi^2} \, 2 \, 
 \int_0^1 \frac{dz}{z(1-z)}
\int_0^\infty \pi dk^2 \psi_V(z,k^2)
 \\
\nonumber
\int_0^\infty
 {\pi d\kappa^2 \over \kappa^4} \alpha_S(q^2) {\cal{F}}(x_{eff},\kappa^2)
\Big( A_L(z,k^2) \; W_L(k^2,\kappa^2)
\Big) \, ,
\end{eqnarray}
where
\begin{eqnarray}
A_L(z,k^2) &=& 4 Q^{2} M z^{2} (1 - z^{2}) \Big (1 +\frac{(1 - 2 z ^{2})}{4 z (1 - z)} \frac{2 m_q}{M + 2 m_q} \Big )  \, ,
\end{eqnarray}
\begin{eqnarray}
W_L(k^2,\kappa^2) &=& 
{1 \over k^2 + \epsilon^2} - {1 \over \sqrt{(k^2-\epsilon^2-\kappa^2)^2 + 4 \epsilon^2 k^2}}
\, . 
\end{eqnarray}

In our calculation we choose the scale of the QCD running coupling constant at: \\
$q^2 = \max \{ \kappa^2, k^2 + m_q^2 \}$.

The full amplitude, at finite $\Delta^2$, for transverse and longitudinal 
polarization can be written as:
\begin{eqnarray}
{\cal M}_{T}(W,\Delta^2,Q^{2}) = (i + \rho) \, \Im m {\cal M}_{T}(W,\Delta^2=0,Q^{2})
\, \exp \Big( \frac{-B(W) \Delta^2}{2} \Big) \, ,
\label{full_amp}
\end{eqnarray}

\vspace{-1.0cm}

\begin{eqnarray}
{\cal M}_{L}(W,\Delta^2,Q^{2}) = (i + \rho) \, \Im m {\cal M}_{L}(W,\Delta^2=0,Q^{2})
\, \exp \Big( \frac{-B(W) \Delta^2}{2} \Big) \, ,
\label{full_amp}
\end{eqnarray}
where $\rho$ is a ratio of real to imaginary part of the amplitude
\begin{eqnarray}
\rho = {\Re e {\cal M} \over \Im m {\cal M}} =  
\tan \Big ( {\pi \over 2} \, { \partial \log \Big( \Im m {\cal M}/W^2 \Big) \over \partial \log W^2 } \Big)
= \tan \Big( {\pi \over 2 } \, \Delta_{\Pom} \Big)
\, .
\end{eqnarray}
Above $B(W)$ is a slope parameter which in general depends on the photon-proton
center-of-mass energy and is parametrized in the present analysis as:
\begin{eqnarray}
B(W) = B_0 + 2 \alpha'_{eff} \log \Big( {W^2 \over W^2_0} \Big) \, ,
\end{eqnarray}
with:$B_{0} = 11$ GeV$^{-2}$, $\alpha'_{eff} = 0.25$ GeV$^{-2}$, $W_0 = 95$ GeV (\cite{H106}).

Assuming exponential dependence of the amplitude on $\Delta^2$
(see Eq.(\ref{full_amp})) the total cross section for transverse and 
longitudinal polarization
can be writen as:
\begin{eqnarray}
\sigma_{L}(\gamma p \to \phi p) = {1 +  \rho^2 \over 16 \pi B(W)} 
\Big| \Im m { {\cal M}_{T}(W,\Delta^2=0,Q^{2}) \over W^2 } \Big|^2 \, ,
\end{eqnarray}
\begin{eqnarray}
\sigma_{T}(\gamma p \to \phi p) = {1 +  \rho^2 \over 16 \pi B(W)} 
\Big| \Im m { {\cal M}_{L}(W,\Delta^2=0,Q^{2}) \over W^2 } \Big|^2 \, .
\end{eqnarray}

The full cross section is a sum of these two components:

\begin{eqnarray}
\sigma_{tot}(\gamma p \to \phi p) =
\sigma_{T}(\gamma p \to \phi p) \, + \epsilon \,
 \sigma_{L}(\gamma p \to \phi p) .
\end{eqnarray}
In HERA kinematics, $\epsilon \approx 1 $.

Let us collect now the main formulas involving the $q \bar q$
light--cone wave-function of the vector meson
The electronic decay width depends on the model of the $\phi$  meson wave
function as:
\begin{eqnarray}
\Gamma (V \to e^+ e^-) = {4 \pi \alpha_{em}^2 c_v \over 3 M_V^3} 
\, 
\cdot g_V^2 
%\cdot K_{NLO} \,
\, .
\end{eqnarray}
In our calculation we use leading-order approximation, i.e. we neglect
a possible NLO $K$-factor. 
The parameter $g_V$ can be expressed in terms of the $\phi$ - meson wave
function as \cite{INS06,Ivanov03}
\begin{eqnarray}
g_V = {8 N_c \over 3} \int {d^3 \vec k \over (2 \pi)^3} 
(M + m_q) \,  \psi_V(k^2) \,
\, .
\end{eqnarray}

Following \cite{INS06,Ivanov03} we use two types of the wave function,
the Gaussian one, representing a standard harmonic--oscillator type
quark model:
\begin{eqnarray}
\psi_{1S}(k^2) = C_1 
\exp\left( - \frac{k^2 a_1^2}{2} \right) \, , \, 
\label{harmonic_oscillator_WF}
\end{eqnarray}
and a Coulomb wave function, representative of models 
in which the wave function has a long high--momentum tail:
\begin{eqnarray}
\psi_{1S}(k^2) = {C_1 \over \sqrt{M}} \, {1 \over (1 + a_1^2 k^2)^2} \, .
\end{eqnarray}
The parameters of the wave function are obtained by fitting to 
the decay width into $e^{+}$ $e^{-}$ and imposing 
the normalization condition:

\begin{eqnarray}
1 = \frac {N_{c} 4 \pi}{(2\pi)^{3}}
\int_0^\infty k^{2}dk\ 4M \psi_{1S}^2(k^2) \, . 
\label{normalization}
\end{eqnarray}

%-------------------------------------------------------------------------------
\section{Exclusive photoproduction of $\phi$ in $p p$ and $p \bar p$ collisions}
%-------------------------------------------------------------------------------

%----------------------------------------------------------------------------------------
\begin{figure}[!h]
\includegraphics[width=0.8\textwidth]{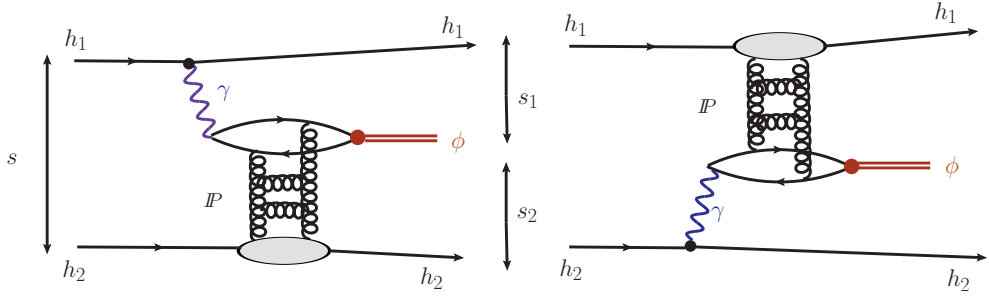}
   \caption{\label{VM_amplitude_wabs}
   \small  A sketch of the bare exclusive $\gamma p \to \phi p$ amplitude.}
\end{figure}
%----------------------------------------------------------------------------------------

%--------------------------------------------------------------------------------------------
\begin{figure}[!h]
\includegraphics[width=0.8\textwidth]{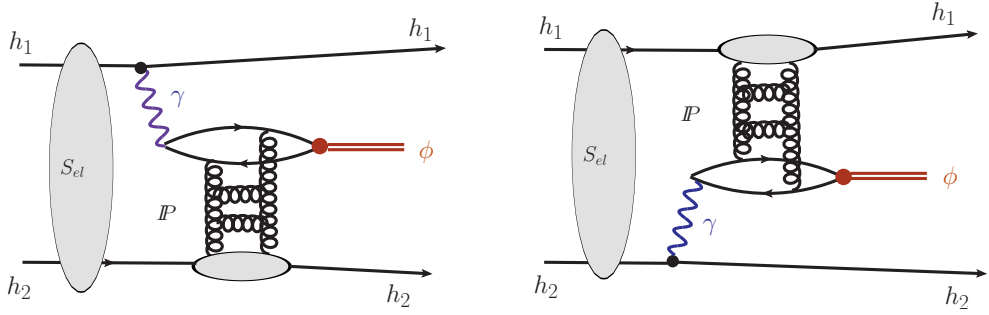}
   \caption{\label{VM_amplitude_abs}
   \small  A sketch of the exclusive $\gamma p \to \phi p$ amplitude with absorptive 
corrections.}
\end{figure}
%--------------------------------------------------------------------------------------------

%\vspace{1.0cm}

The bare amplitude (when absorption effects are ignored) can be
written schematically \cite{SS07} as:
\begin{eqnarray}
\bM^{(0)}(\bp_1,\bp_2) &&= e_1 {2 \over z_1} {\bp_1 \over t_1} 
{\cal{F}}_{\lambda_1' \lambda_1}(\bp_1,t_1)
{\cal {M}}_{\gamma^* h_2 \to V h_2}(s_2,t_2,Q_1^2)
\\
&&
+ e_2 {2 \over z_2} {\bp_2 \over t_2} {\cal{F}}_{\lambda_2' \lambda_2}(\bp_2,t_2)
{\cal {M}}_{\gamma^* h_1 \to V h_1}(s_1,t_1,Q_2^2)
\, .
\nonumber
\end{eqnarray}
Because of the presence of the proton from factors only small $Q_{1}^{2}$ and $Q_{2}^{2}$
enter the amplitude for the hadronic process. This means that in practise one can put
$Q_{1}^{2} = Q_{1}^{2} = 0 $

The full amplitude for the $p p$ $\longrightarrow$ $p \phi p$, reaction which includes
elastic rescatterings reads:
\begin{eqnarray}
\bM(\bp_1,\bp_2) &&= \int{d^2 \bk \over (2 \pi)^2} \,S_{el}(\bk) \,
\bM^{(0)}(\bp_1 - \bk, \bp_2 + \bk)\\
&&= \bM^{(0)}(\bp_1,\bp_2) - \delta \bM(\bp_1,\bp_2)\, ,
\nonumber 
\end{eqnarray}
where
\begin{equation}
S_{el}(\bk) = (2 \pi)^2 \delta^{(2)}(\bk) - \half T(\bk) \, \, \, ,
\, \, \, T(\bk) = \sigma^{pp}_{tot}(s) \, \exp \Big(-\half B_{el} \bk^2 \Big) \, .
\end{equation}
In practical evaluations we take $B_{el} = 14$ GeV$^{-2}$, \ $\sigma^{pp}_{tot} = 52$ mb
for the RHIC energy $W$ = 200 GeV, $B_{el} = 17$ GeV$^{-2}$, \ $\sigma^{pp}_{tot} = 76$ mb
\cite {CDF94} for the Tevatron energy $W$ = 1.96 TeV and $B_{el} = 21$ GeV$^{-2}$,
\ $\sigma^{pp}_{tot} = 100$ mb for the LHC energy $W$ = 14 TeV.

\vspace{0.3cm}

The absorptive correction to the amplitude can be written as the following convolution
of the bare amplitude $(M^{(0)})$ and the amplitude for elastic $pp$ or $p \bar p$
scattering as:
\begin{eqnarray}
\delta \bM(\bp_1,\bp_2) = \int {d^2\bk \over 2 (2\pi)^2} \, T(\bk) \,
\bM^{(0)}(\bp_1-\bk,\bp_2+\bk) \, .
\end{eqnarray}

The differential cross section is expressed in terms of the amplitude $\bM$ as
\begin{eqnarray}
d \sigma = { 1 \over 512 \pi^4 s^2 } | \bM |^2 \, dy dt_1 dt_2
d\varphi
\, .
\end{eqnarray}
This formula is used below to calculate several differential distributions
in $p p \to p \phi p$ and $p \bar p \to p \phi \bar p$ reactions.

\vspace{1.0cm}

%----------------
\section{Results}
%----------------

%------------------------------------------------------------------------------------
\begin{figure}[!htb]
\begin{center}
\includegraphics[height=4.4cm]{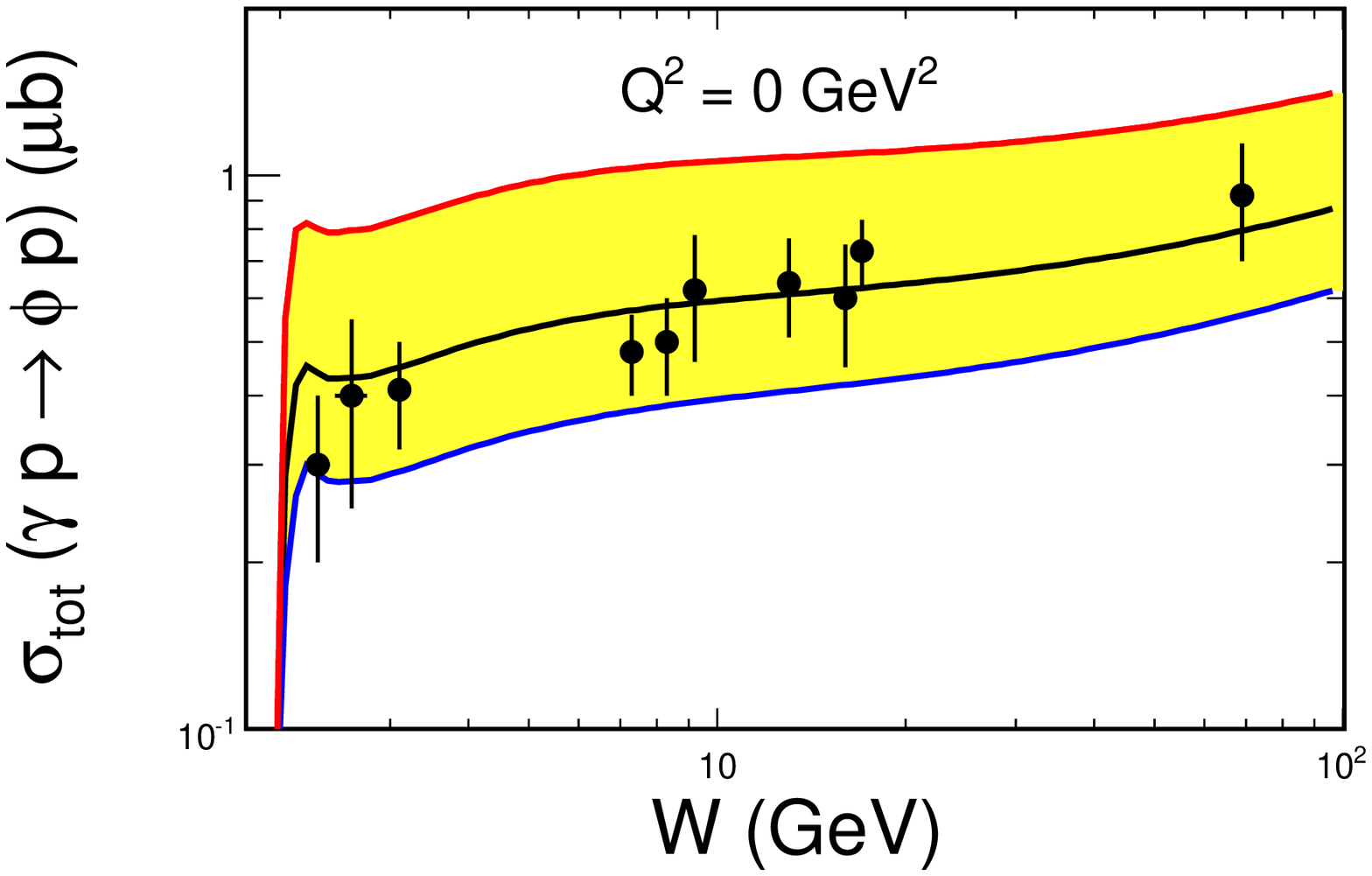}
\vspace{0.35cm}
\includegraphics[height=4.8cm]{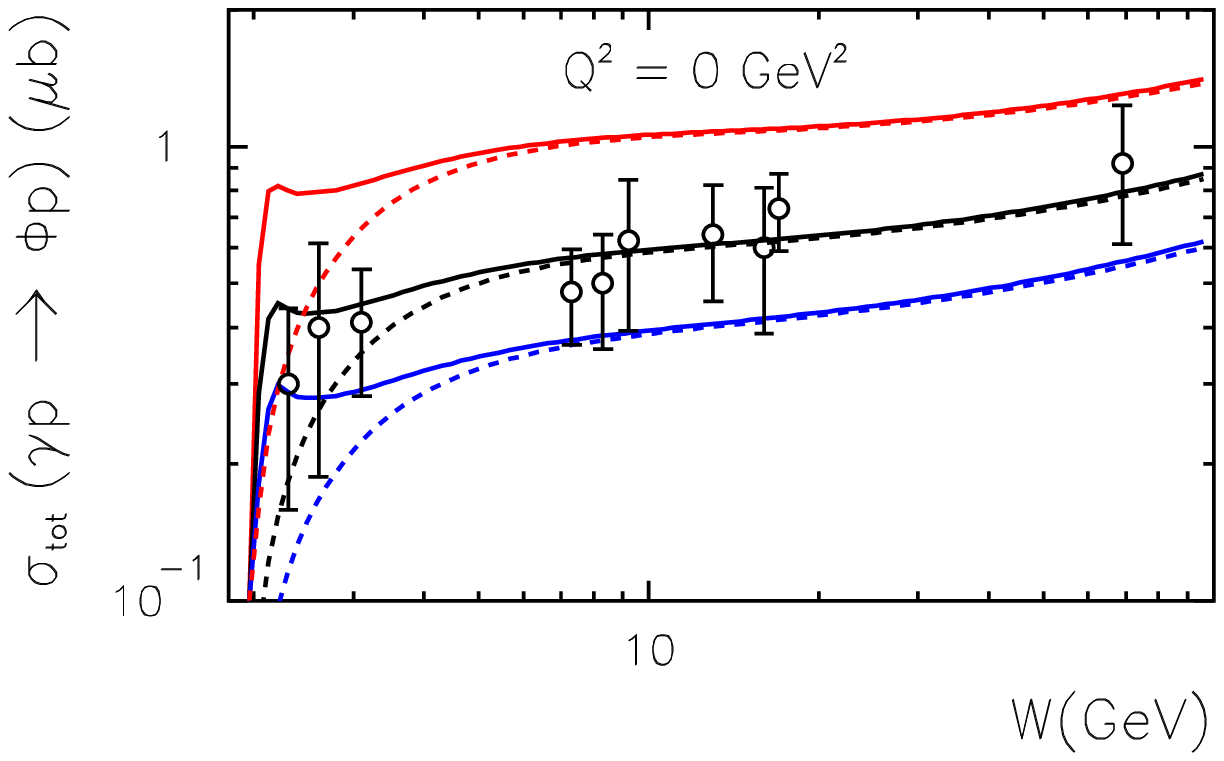}
\caption[*]{Total cross section for the photoproduction $\gamma p \to \phi p$ process 
as a function of the photon-proton center-of-mass energy. In this calculation
the Gaussian wave function is used. The curves are described in the text.}
\label{sigma_tot_mass}
\end{center}
\end{figure}
%-------------------------------------------------------------------------------------

Let us start review of our results with the $\gamma p \to \phi p$ reaction.
In Fig.\ref {sigma_tot_mass} we show the total cross section for
$\gamma p \to \phi p$ as a function of photon-proton center of-mass energy $W$ for
$Q^{2} = 0 \ GeV^2$.
Our results are compared with the corresponding HERA data \cite{ZEUS96}.
In order to describe experimental data we treat the mass of the strange
quark as a free parameter, although it was fixed at $m_{S}$ = 0.37 GeV in
Ref. \cite{IN02}.
We show results for three different values of the strange quark mass.
The red solid (upper) line is for $m_{S}$ = 0.37 GeV, blue (lower) line for
$m_{S}$ = 0.50 GeV and the black line (which goes through the data points)
for $m_{S}$ = 0.45 GeV. We can see that the result for $m_{S}$ = 0.45 GeV
gives the best description of the experimental data,
so below we shall present results only for this value of the strange quark mass.
Our results for $Q^{2}$-dependence look consistent with the ones previously
obtained by Ivanov in his phD thesis.\cite{Ivanov03}

%-----------------------------------------------------------------------------------------
\begin{figure}[!htb]
\begin{center}
\includegraphics[height=4.8cm]{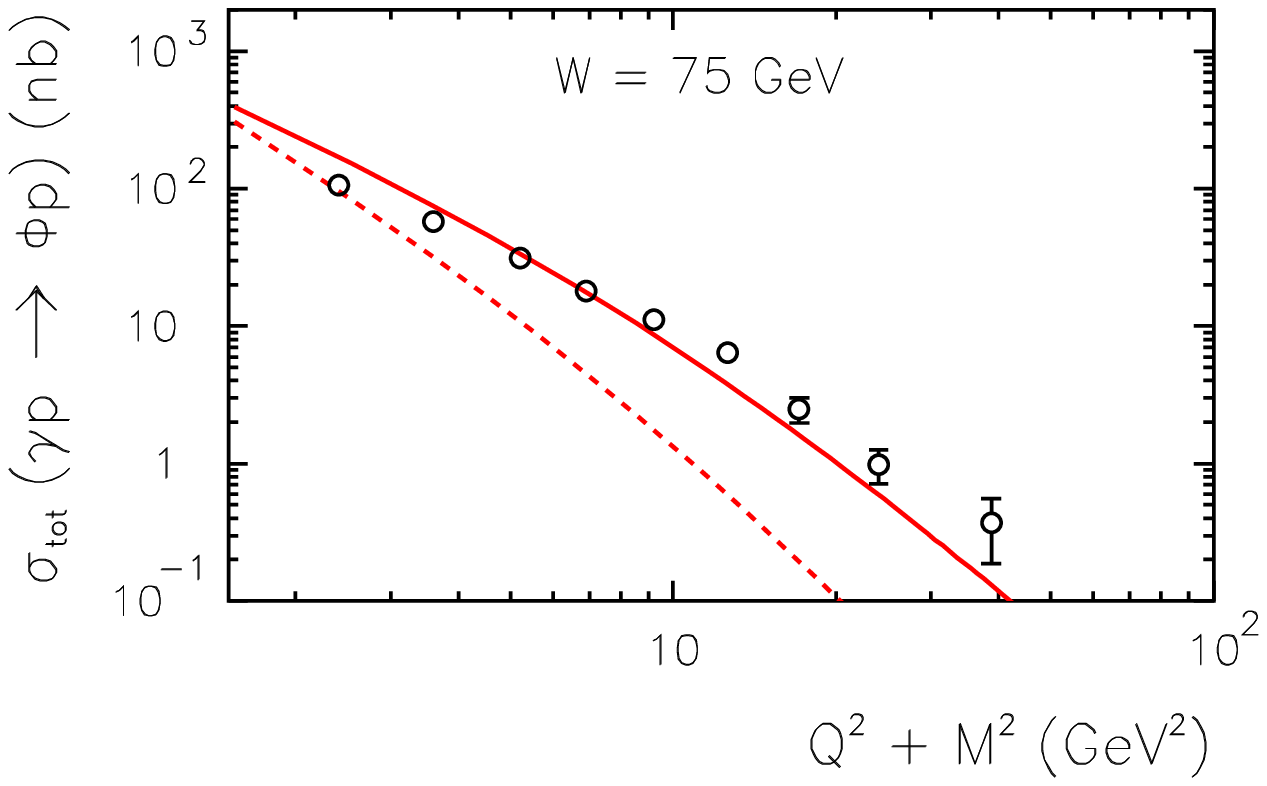}
\includegraphics[height=4.8cm]{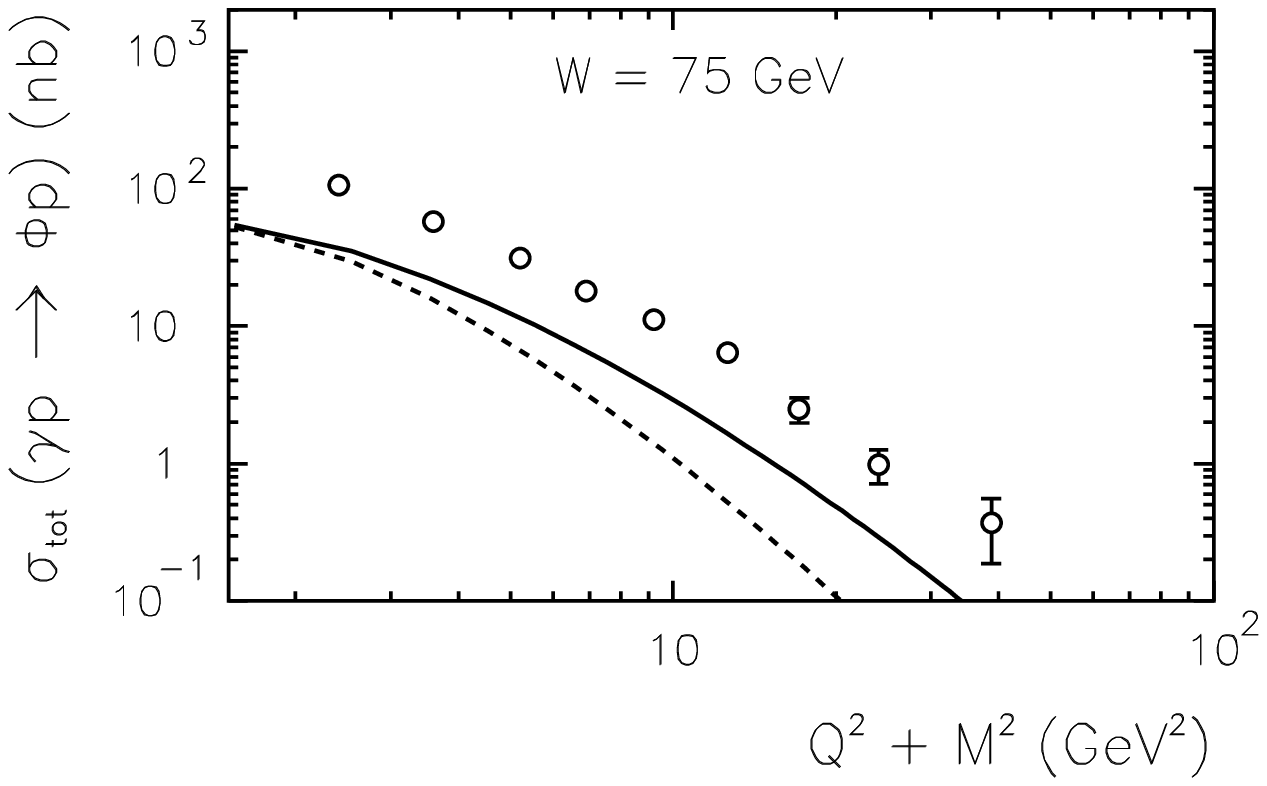}
\caption[*]{Total cross section as a function of photon virtuality for two different
wave functions: Gauss (left) and Coulomb (right) panel respectively. Here $m_s$ = 0.45 GeV.
The solid line is for transverse and longitudinal cross section and the dashed line is
only for transverse cross section.}
\label{sigma_tot_Q2}
\end{center}
\end{figure}
%-----------------------------------------------------------------------------------------

In Fig.\ref{sigma_tot_Q2} we show the total cross section as a function of 
photon virtuality for photon-proton center-of-mass energy $W$ = 75 GeV. 
The data points are taken from Ref.\cite{ZEUS05}. 
The solid line is for the sum of longitudinal and transverse cross sections 
($\sigma_{L}$ + $\sigma_{T}$) and the dashed line is for
transverse cross section ($\sigma_{T}$) alone. 
It is important that both these components are included in the calculation.
In the left panel we show results for the Gaussian wave function 
and in the right panel for the Coulomb wave function.
We can see that the Gaussian wave function much better describes 
the experimental data than the Coulomb wave function.

%----------------------------------------------------------------------------------------
\begin{figure}[!htb]
\begin{center}
\includegraphics[height=4.8cm]{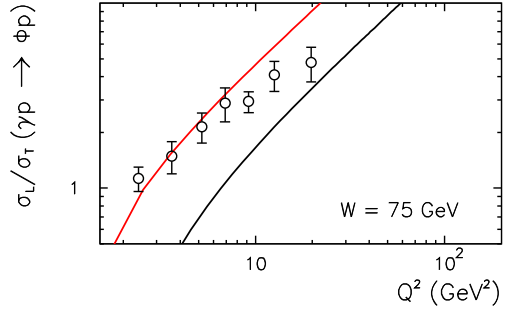}
\caption[*]{The ratio $\sigma_{L}$/$\sigma_{T}$ as a function of photon virtuality. 
The red solid line is for the Gaussian wave function and the black solid line is for 
the Coulomb wave function. Here $m_s$ = 0.45 GeV.}
\label{ratio}
\end{center}
\end{figure}
%------------------------------------------------------------------------------------------

In Fig.\ref{ratio} we show the ratio of the longitudinal cross section 
($\sigma_{L}$) to the transverse cross section ($\sigma_{T}$) as a function 
of photon virtuality, again for $W$ = 75 GeV. The red solid line is for 
the Gaussian and the black line is for the Coulomb wave function.
Our results are compared with the HERA data from Ref.\cite{ZEUS05}.

%--------------------------------------------------------------------------------------------
\begin{figure}[!htb]
\begin{center}
\includegraphics[height=3.6cm]{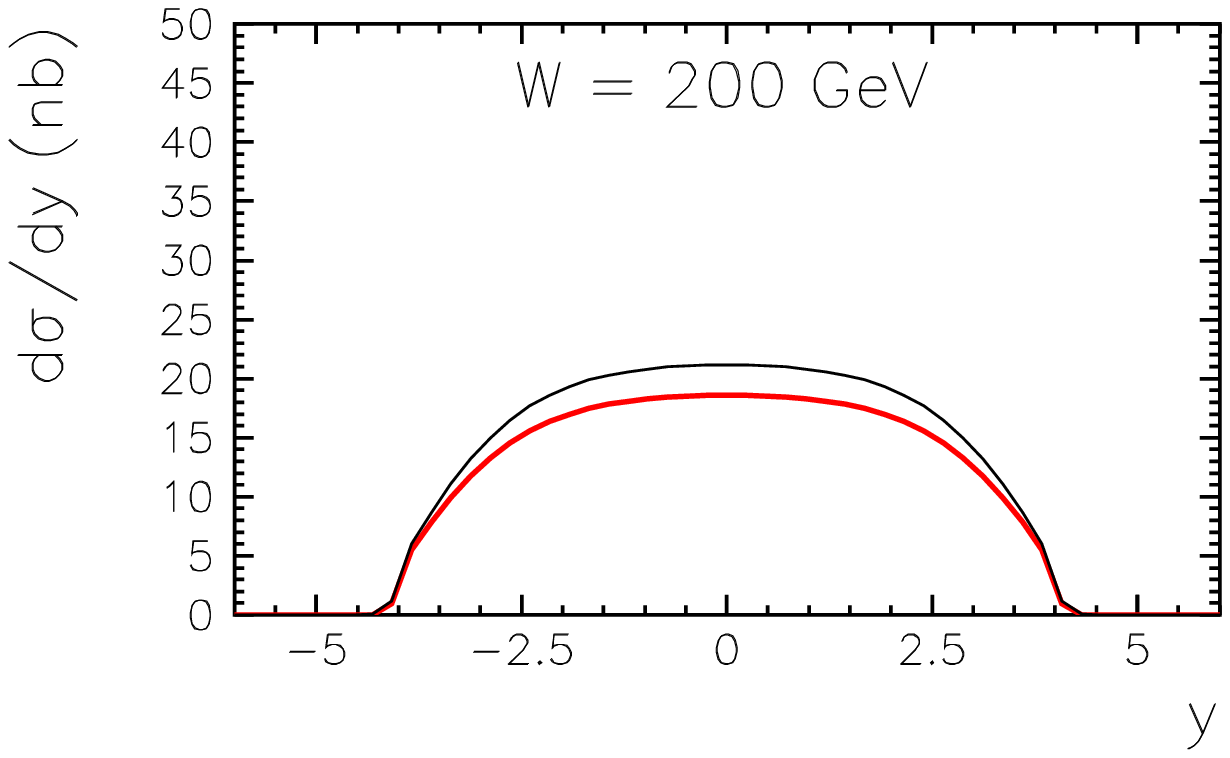}
\includegraphics[height=3.6cm]{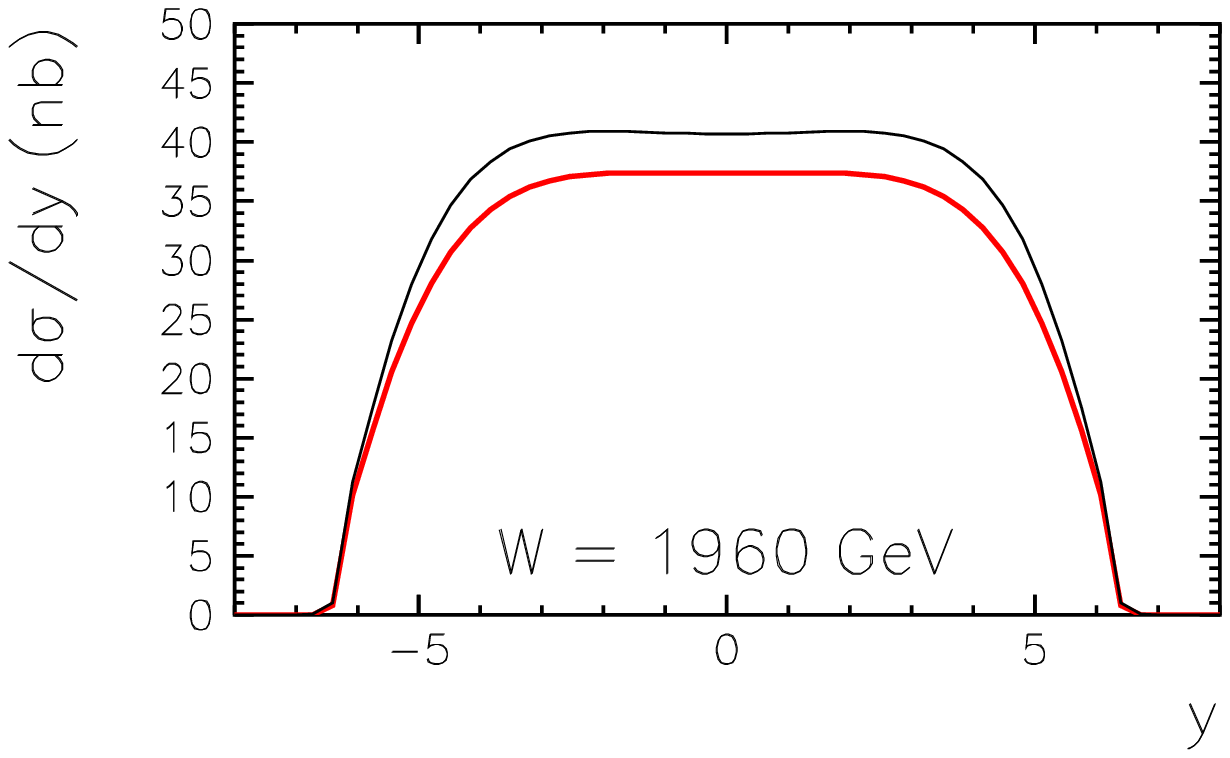}
\includegraphics[height=3.6cm]{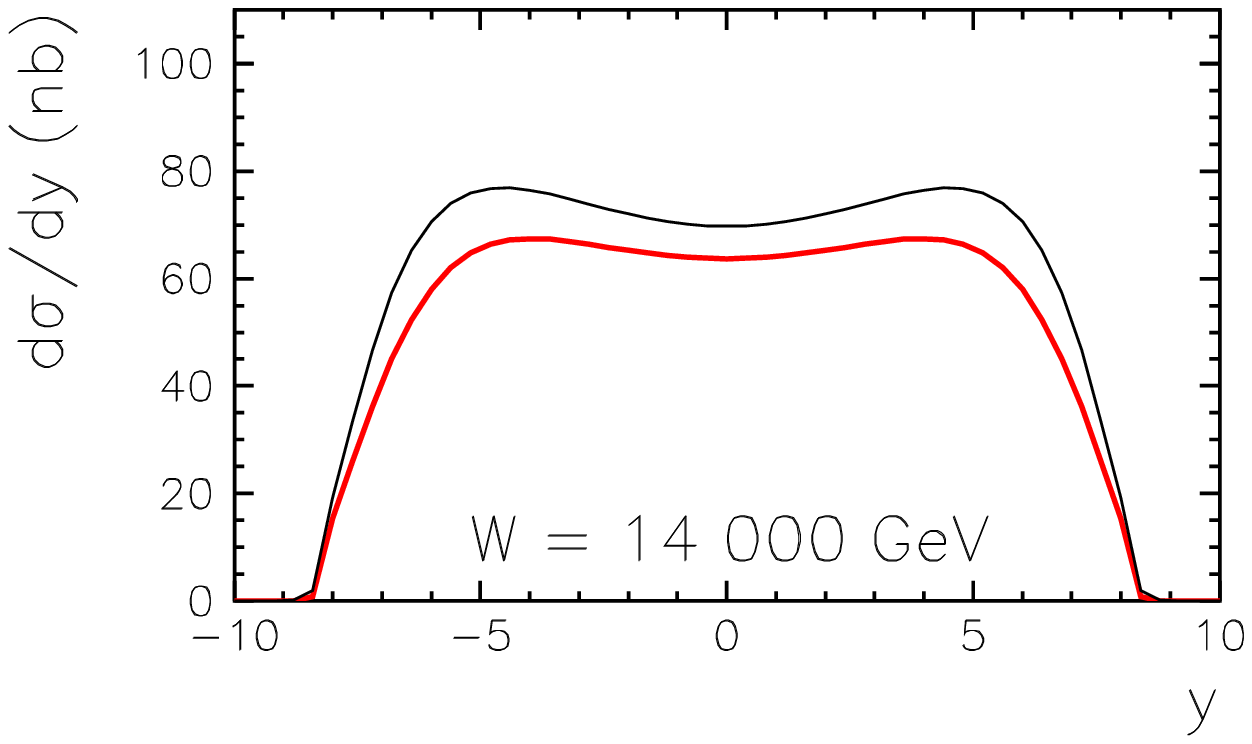}
\caption[*]{Rapidity spectrum of exclusive $\phi$  production at RHIC (left), Tevatron (middle)
and LHC (right) energies. The thin black lines are without, the thick red lines with absorption
included. In this calculation Gaussian wave function was used and $m_s$ = 0.45 GeV.}
\label{rapidity_absorption}
\end{center}
\end{figure}
%--------------------------------------------------------------------------------------------

Let us come now to the presentation of our results for exclusive production
of $\phi$ meson in hadronic reactions.
In Fig.\ref{rapidity_absorption} we show distributions in rapidity for the
$p \bar p \to p \phi \bar p $ (Tevatron) and $p p \to p \phi p$ (RHIC, LHC)
reactions without (black thin solid) and with (grey thick solid) absorption effects.
The absorption effects are rather small and will be not included in calculations of
other observables.

%-----------------------------------------------------------------------------------------
\begin{figure}[!htb]
\begin{center}
\includegraphics[height=3.6cm]{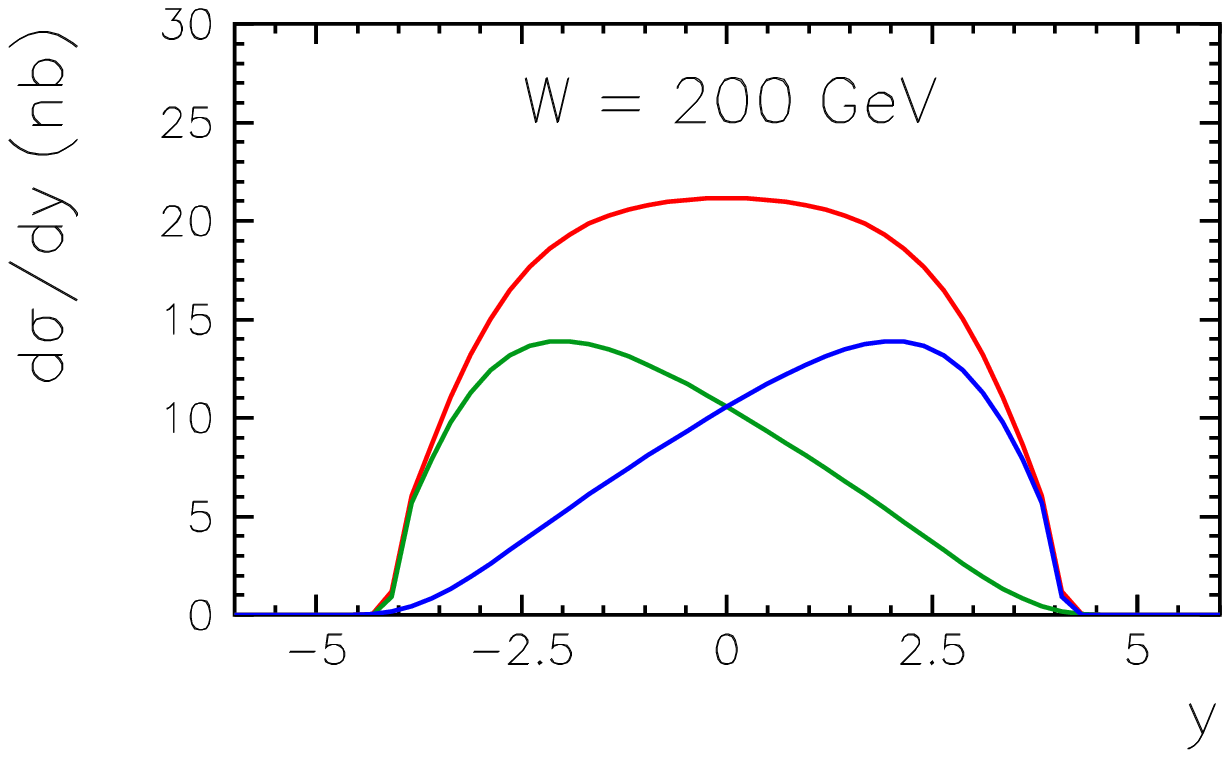}
\includegraphics[height=3.6cm]{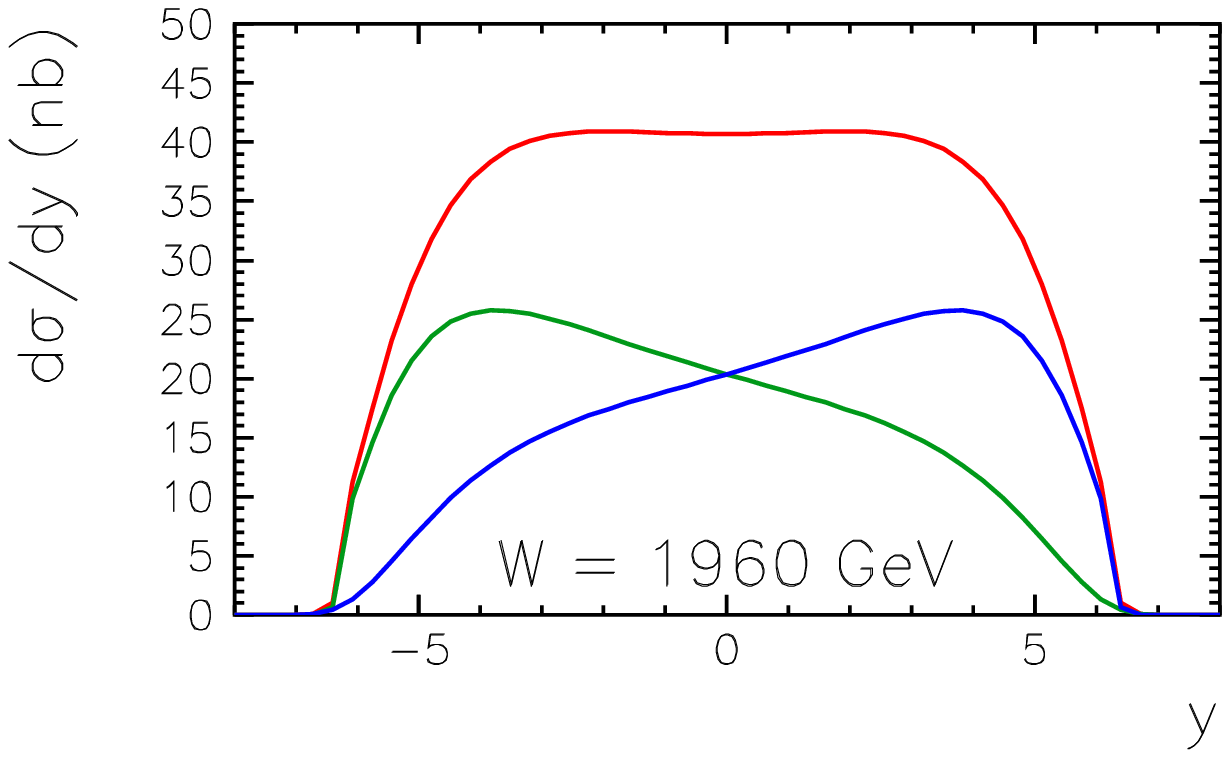}
\includegraphics[height=3.6cm]{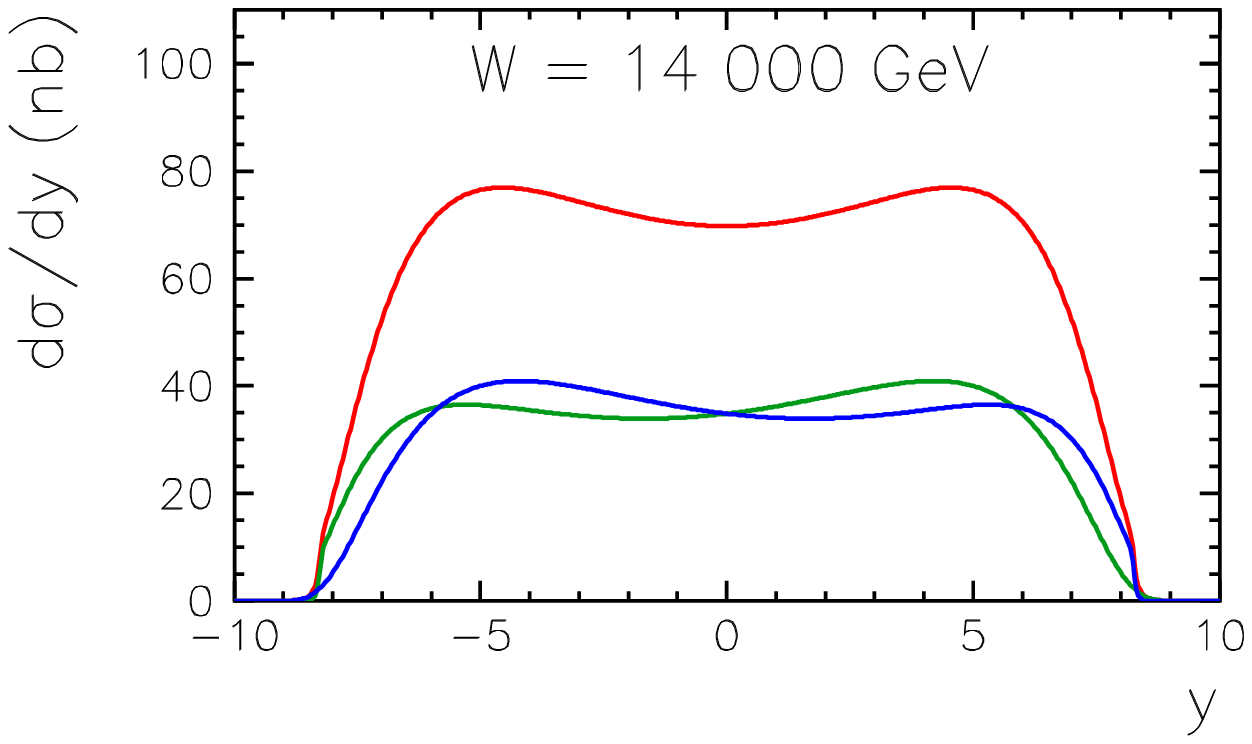}
\caption[*]{The photon-pomeron and pomeron-photon contributions to the rapidity
distribution for the RHIC (left), Tevatron (middle) and LHC (right) energies. Here Gaussian wave
function was used, $m_s$ = 0.45 GeV and absorptive corrections are not included.}
\label{rapidity_gppg}
\end{center}
\end{figure}
%-------------------------------------------------------------------------------------------

In Fig.\ref{rapidity_gppg} we show separate contributions of photon-pomeron and 
pomeron-photon fusion mechanisms in rapidity distribution of $\phi$ in $p \bar p$
(middle panel) and $p p$ (left and right panels) collisions. Here we have shown only
results without absorptive corrections. The green line is for photon-pomeron,
the blue line is for pomeron-photon fusion mechanism and the red most upper
line is a sum of the two components. In rapidity distribution the two contributions
add incoherently \cite{SS07}.

%--------------------------------------------------------------------------------------
\begin{figure}[!htb]
\begin{center}
\includegraphics[height=3.6cm]{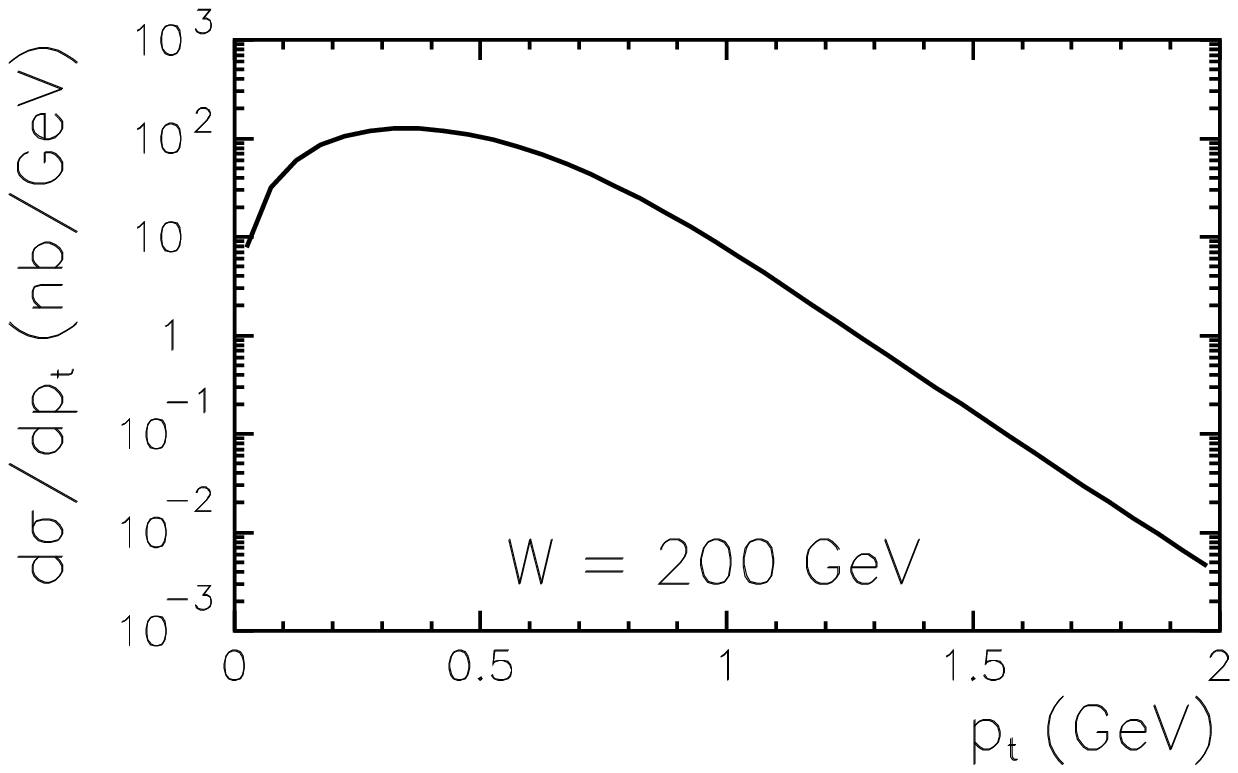}
\includegraphics[height=3.6cm]{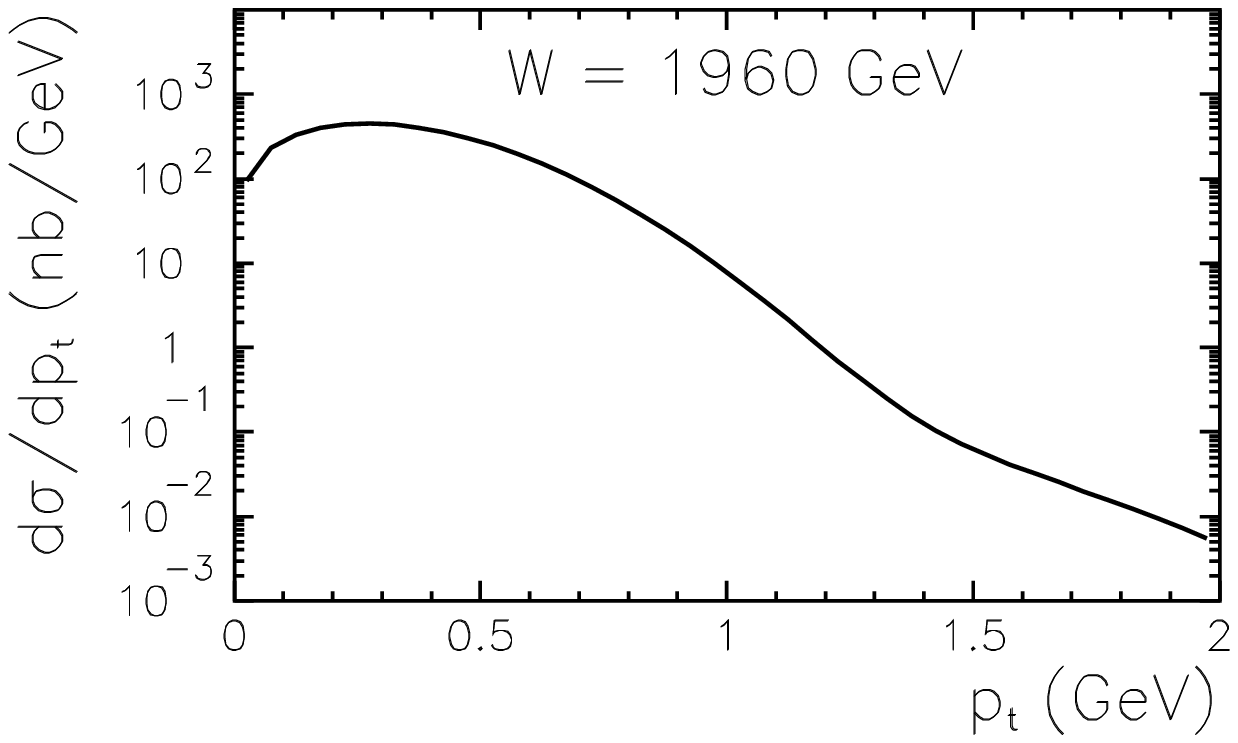}
\includegraphics[height=3.6cm]{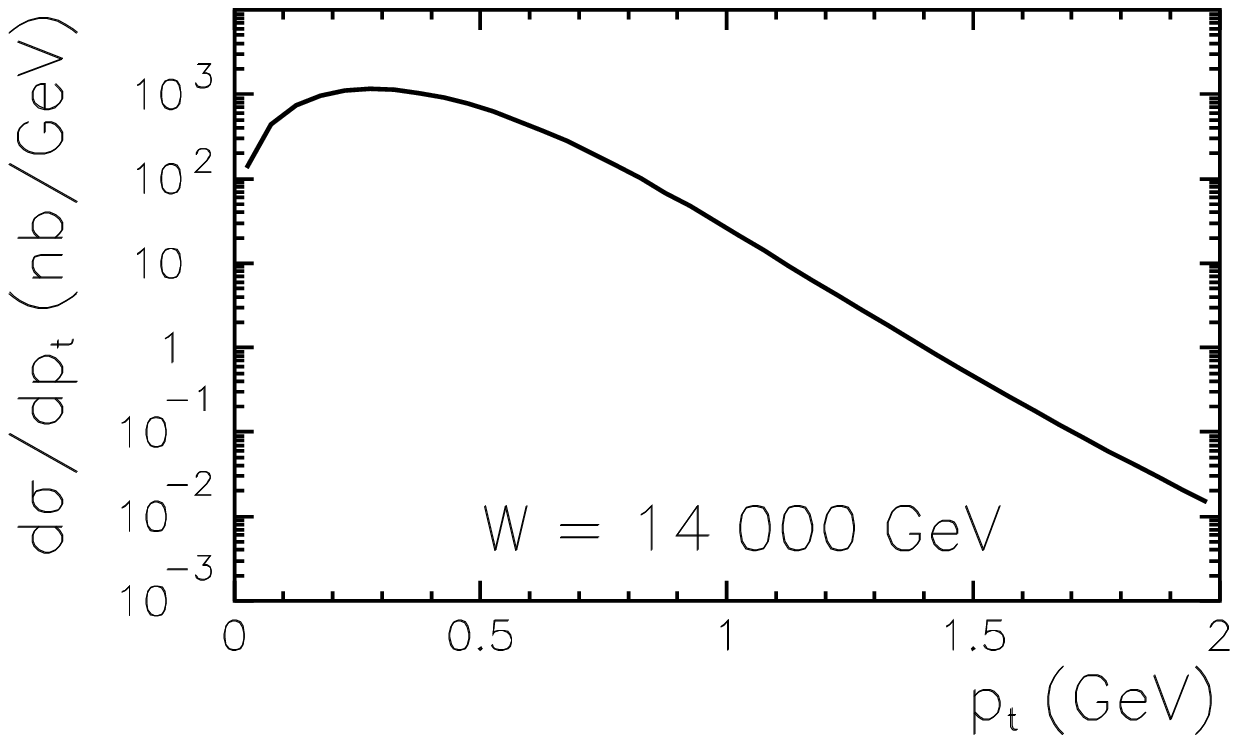}
\caption[*]{The transverse momentum distribution for the RHIC (left), Tevatron (middle)
and LHC (right) energies.}
\label{transverse_momentum}
\end{center}
\end{figure}
%---------------------------------------------------------------------------------------

In Fig.\ref{transverse_momentum} we show distribution in transverse momentum of 
the exclusively produced meson $\phi$. In the left panel we present results for
RHIC energy, in the middle panel for Tevatron energy and in the right panel for the LHC energy.

%--------------------------------------------------------------------------------------
\begin{figure}[!htb]
\begin{center}
\includegraphics[height=3.6cm]{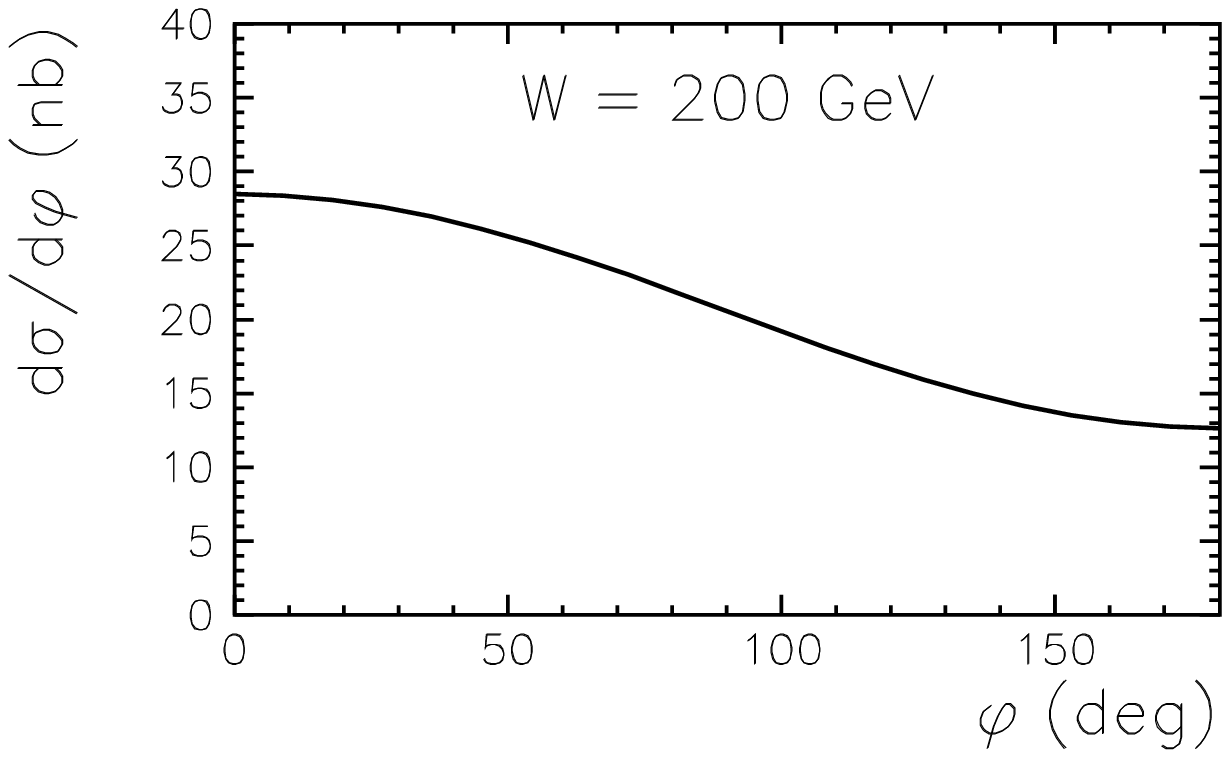}
\includegraphics[height=3.6cm]{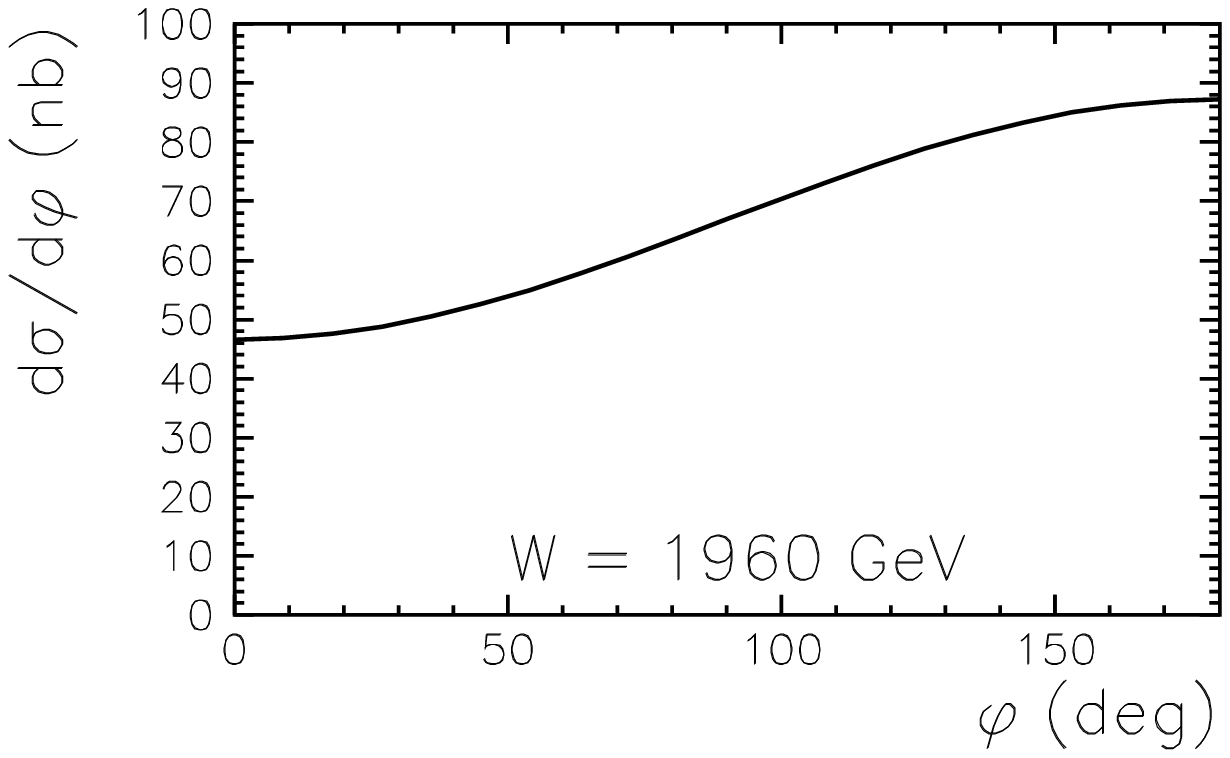}
\includegraphics[height=3.6cm]{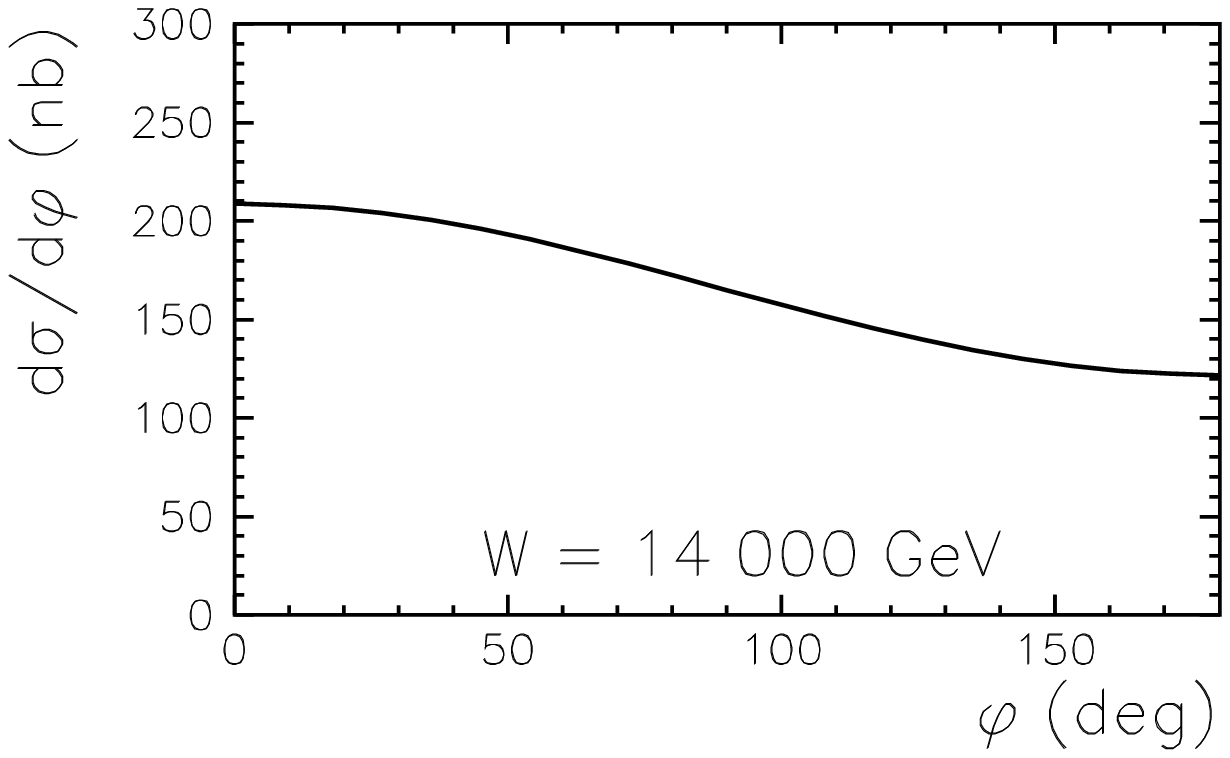}
\caption[*]{Distribution in azimuthal angle  for the RHIC (left), Tevatron (middle)
and LHC (right) energies.}
\label{angle}
\end{center}
\end{figure}
%----------------------------------------------------------------------------------------

Finally in Fig.\ref{angle} we show the distribution in relative 
azimuthal angle between outgoing protons. 
Quite different distributions are obtained for the RHIC, LHC ($pp$ collisions)
and Tevatron ($\bar p p$ collisions). This effect is of the interference nature
and was already discussed for the exclusive production of $J/\Psi$ \cite{SS07}.

%--------------------
\section{Conclusions}
%--------------------

In the present analysis we have performed the calculation of the cross
section for the $\gamma p \to \phi p$ process as well as for
exclusive production of $\phi$ meson in the $p p \to p \phi p$
reaction.

The cross section for the first process depends strongly
on the model of the $\phi$ meson wave function as well as on the strange quark
mass. Both Gaussian and Coulomb wave functions were used with
parameters adjusted to reproduce the electronic decay width of $\phi$.
The strange quark mass was adjusted to reproduce the 
$\gamma p \to \phi p$ cross section measured by the ZEUS collaboration
at HERA in the Gaussian model of the wave function.

So-fixed parameters were used to make predictions for the
$p \bar p \to p \bar p \phi$ and $p p \to p \phi p$ reactions.
We have presented distributions in the $\phi$ meson rapidity,
in the $\phi$ meson transverse momentum and in the relative azimuthal 
angle between outgoing protons. 
The azimuthal angle dependence found here is of purely
interference nature. Quite different distributions have been
predicted for the Tevatron and LHC due to the different sign of 
the interference term (different electric charge of proton 
and antiproton).  

The cross sections found in the present analysis are fairly large.
In practice one could measure the $\phi$ meson either in 
the lepton-antilepton or in kaon-antikaon decay channels. 
The later method could be used by the ALICE collaboration at the LHC
where even small transverse momenta of charged kaons can be registered
\cite{ALICE}. 
A feasibility of the measurements requires, however, detailed Monte Carlo 
studies.

%-----------------------------------------------------------------------
\vspace{0.7cm}

{\bf Acknowledgments}
We are indebted to Aharon Levy for providing us with the ZEUS 
experimental data.
This work was partially supported by the Polish grants of MNiSW 
No. N202 249235 and No. N202 191634 .

%
%======================================================================

\end{document}